\documentstyle[12pt,epsf]{article}
\def \apk{A_{\pi K}}
\def \abpk{\overline{A}_{\pi K}}
\def \app{A_{\pi \pi}}
\def \abpp{\overline{A}_{\pi \pi}}
\def \beq{\begin{equation}}
\def \cap{{\cal P}}
\def \capp{{\cal P'}}
\def \cat{{\cal T}}
\def \catp{{\cal T'}}
\def \eeq{\end{equation}}

\def \st{\sqrt{3}}

\def \tcap{\tilde{\cal P}}
\def \tcapp{\tilde{\cal P'}}
\def \tru{\tilde r_u}
\begin{document}
\renewcommand{\thetable}{\Roman{table}}
\rightline{CERN-TH/96-68}
\rightline{EFI-96-10}
\rightline{TECHNION-PH-96-05}
\rightline{hep-ph/9604233}
\rightline{April 1996}
\bigskip
\bigskip
\centerline{{\bf WEAK PHASES FROM $B$ DECAYS TO KAONS AND CHARGED PIONS}
\footnote{To be submitted to Phys.~Rev.~D.}}
\bigskip
\centerline{\it Amol S. Dighe}
\centerline{\it Enrico Fermi Institute and Department of Physics}
\centerline{\it University of Chicago, Chicago, IL 60637}
\medskip
\centerline{\it Michael Gronau}
\centerline{\it Department of Physics}
\centerline{\it Technion -- Israel Institute of Technology, Haifa 32000,
Israel}
\medskip
\centerline{and}
\medskip
\centerline{\it Jonathan L. Rosner}
\medskip
\centerline{\it Div.~TH, CERN}
\centerline{\it 1211 CH Geneva 23, Switzerland}
\smallskip
\centerline{and}
\smallskip
\centerline{\it Enrico Fermi Institute and Department of Physics}
\centerline{\it University of Chicago, Chicago, IL 60637
\footnote{Permanent address.}}
\bigskip
\centerline{\bf ABSTRACT}
\medskip
\begin{quote}

Phases of elements of the Cabibbo-Kobayashi-Maskawa (CKM) matrix can be
obtained using decays of $B$ mesons to $\pi^+ \pi^-$, $\pi^\pm K^\mp$, and
$\pi^+ K^0$ or $\pi^- \overline{K}^0$. For $B^0~{\rm or}~ \overline{B}^0 \to
\pi^+ \pi^-$, one identifies the flavor of the neutral $B$ meson at time of
production and studies the time-dependence of the decay rate. The other
processes are self-tagging and only their rates need be measured. By assuming
flavor SU(3) symmetry and first-order SU(3) breaking, one can separately
determine the phases $\gamma \equiv {\rm Arg}~V_{ub}^*$ and $\alpha = \pi -
\beta - \gamma$, where $\beta \equiv {\rm Arg}~V_{td}^*$. Special cases include
the vanishing of strong interaction phase differences between amplitudes, the
possibility of recovering partial information when $\pi^+ \pi^-$ and $\pi^\pm
K^\mp$ decays cannot be distinguished from one another, and the use of a
correlation between $\gamma$ and $\alpha$ in the region of allowed parameters. 
\end{quote}
\leftline{PACS codes:  11.30.Er, 12.15.Hh, 13.25.Hw, 14.40.Nd}
\medskip
\leftline{CERN-TH/96-68}
\leftline{April 1996}
\newpage

\centerline{\bf I.  INTRODUCTION}
\bigskip

The decays of $B$ mesons \cite{CPreview} offer the prospect of confirming or
refuting the current explanation of CP violation in the neutral kaon system
\cite{CCFT}, based on phases in the Cabibbo-Kobayashi-Maskawa (CKM) matrix
\cite{CKM}.  For example, unequal time-integrated rates for the $\pi^+ \pi^-$
decays of states which are initially $B^0$ and and $\overline{B}^0$ would
signify CP violation, providing approximate information on the angle $\alpha$
of the triangle describing the unitarity of the CKM matrix. 

The presence of gluonic \cite{Penguin} and electroweak \cite{EWP} penguin
contributions in additional to the dominant (``tree'') processes requires that
one separate out several terms. An isospin analysis \cite{Isospin}, involving
the study of the time dependence of the $\pi^+ \pi^-$ mode and rates for the
$\pi^\pm \pi^0$ and $\pi^0 \pi^0$ modes of $B$ mesons, permits one to isolate
the amplitudes contributing to final states with isospin $0$ and $2$ and
thereby to determine $\alpha$ rather well \cite{Isospin,EWPin}.  However, for
certain types of detectors, the observation of neutral pions may pose a
challenge, and model calculations \cite{models} predict a branching ratio for
$B^0 \to \pi^0 \pi^0$ of order $10^{-6}$ or less. 

A few alternative ways to sort out the effects of several amplitudes in
$B^0\to\pi^+\pi^-$ were suggested recently.  DeJongh and Sphicas \cite{DS}
studied the dependence of the asymmetry in $B^0(t)\to\pi^+\pi^-$ on the
magnitude and relative phase of the contributing terms. Using flavor SU(3)
symmetry, Silva and Wolfenstein \cite{SilWo} estimated the penguin contribution
by comparing the tree-dominated decay rate of $B^0\to\pi^+\pi^-$ with that of
$B^0\to\pi^- K^+$ which has a large penguin term. Buras and Fleischer
\cite{BurFl} proposed relating the penguin term in $B^0\to\pi^+\pi^-$ via SU(3)
to the time-dependent asymmetry of $B^0(t)\to K^0\overline{K}^0$, where the
penguin amplitude dominates.  Kramer, Palmer, and Wu \cite{KPW} note that the
ratio of penguin to tree matrix elements is less model-dependent than either
quantity alone, and thereby obtain a relation for $\alpha$.  Aleksan {\it et
al.} \cite{Aleksan} use model-dependent assumptions to learn the magnitude of
the penguin effect on the measurement of $\alpha$ by relating the three $\Delta
S = 0$ decay modes $\pi \pi$, $\pi \rho$, and $\rho \rho$ to the corresponding
$\Delta S = 1$ modes $\pi K$, $\pi K^*$ and $\rho K$, and $\rho K^*$. 

In this paper we examine in more detail a method proposed in Ref.~\cite{GRPRL}
to determine phases of CKM matrix elements by detecting only kaons and charged
pions in $B$ meson decays. In the decays $B^0 \to \pi^+ \pi^-$ and
$\overline{B}^0$ to $\pi^+ \pi^-$, one identifies the flavor of the neutral $B$
meson at time of production and studies the time-dependence of the decay rate.
One obtains the necessary information on additional amplitudes from the rates
$\Gamma(B^0 \to \pi^- K^+)$, $\Gamma(\overline{B}^0 \to \pi^+ K^-)$, and
$\Gamma(B^+ \to \pi^+ K^0)$ or $\Gamma(B^- \to \pi^+ \overline{K}^0)$ using
flavor SU(3) symmetry \cite{SilWo,DZ,SW,Chau,SU3} and first-order SU(3)
breaking \cite{SU3br}.  In the most general case we obtain information not
only on $\alpha$, but also on $\gamma = {\rm Arg}(V_{ub}^*)$ and on strong
phase shift differences.  Other ways to measure $\gamma$, based on charged $B$
decays, were proposed in Ref.~\cite{gamma}. 

In Section II we describe the processes to be measured and the amplitudes on
which they depend.  We then study the precision to which various quantities can
be determined. It is possible that the strong-interaction phase shift
difference $\delta$ between amplitudes is below detectable levels, in which
case simplified analyses become necessary.  Several of these cases are
discussed in Sec. III.  The most general error analysis (for $\delta \ne 0$) is
performed in Sec.~IV, while Sec.~V concludes.  An Appendix is devoted to an
aspect of Monte Carlo programs. 
\bigskip

\centerline{\bf II.  PROCESSES AND AMPLITUDES}
\bigskip

\leftline{\bf A.  Expressions for amplitudes and quantities quadratic in them}
\bigskip

We review the method proposed in Ref.~\cite{GRPRL}, which may be consulted for
details. Our method employs flavor SU(3) symmetry \cite{DZ,SW,Chau}, and
neglects ``annihilation'' amplitudes in which the spectator quark (the light
quark accompanying the $b$ in the initial meson) enters into the decay
Hamiltonian \cite{SU3}. These amplitudes in $B$ decays are expected to be
suppressed by $f_B/m_B$, where $f_B \simeq 180$ MeV.  We include first-order
SU(3) breaking terms \cite{SU3br}, expected to be at most tens of percent, but
neglect corrections expected to arise at a level of a few percent. 

In the SU(3) limit and neglecting annihilation terms, all $B$ decay amplitudes
into $\pi\pi~,\pi K$ and $K\overline{K}$ states can be decomposed in terms of
three independent amplitudes \cite{EWPin,SU3}:  a ``tree'' term $t(t')$, a
``color-suppressed'' term $c(c')$ and a ``penguin'' term $p(p')$. These
amplitudes contain both the leading-order and electroweak penguin \cite{EWP}
contributions: 
\beq
t \equiv T + (c_u - c_d) P_{EW}^C~~,~~~c \equiv C + (c_u - c_d) P_{EW}~~,
~~~p \equiv P + c_d P_{EW}^C~~.
\eeq
Here the capital letters denote the leading-order contributions defined in
Ref.~\cite{SU3}, and $P_{EW}$ and $P_{EW}^C$ are color-favored and
color-suppressed electroweak penguin amplitudes defined in Ref.~\cite{EWPin}.
The values $c_u = 2/3$ and $c_d = -1/3$ are those which would follow if the
electroweak penguin coupled to quarks in a manner proportional to their
charges.  (Small corrections, which we shall ignore and which do not affect our
analysis, arise from axial-vector $Z$ couplings and from $WW$ box diagrams.) 
The $\Delta S = 0$ amplitudes are denoted by unprimed quantities and the
$\Delta S = 1$ processes by primed quantities. 

The amplitudes of the two processes $B^0\to\pi^+\pi^-$ and $B^0\to\pi^- K^+$
are expressed as 
$$
\app \equiv A(B^0\to\pi^+\pi^-)=-t-p=-T-P-{2\over 3}P^C_{EW}~~,
$$
\beq \label{eqn:amps}
\apk \equiv A(B^0\to\pi^- K^+)=-t'-p'=-T'-P'-{2\over 3}P'^C_{EW}~~,
\eeq
while that for $B^+\to \pi^+ K^0$ will be approximated by
\beq \label{eqn:approx}
A_+\equiv A(B^+\to\pi^+ K^0)= p' = P'-{1\over 3}P'^C_{EW}
\approx P' + {2\over 3}P'^C_{EW}~~,
\eeq
neglecting a color-suppressed electroweak penguin effect of order
$|P'^C_{EW}/P'|={\cal O}((1/5)^2)$ \cite{EWPin}. With this approximation,
$A_+$ contains the same combination of electroweak and gluonic penguins
as in the expression for $A_{\pi K}$.

The terms on the right-hand-sides of (\ref{eqn:amps}) and (\ref{eqn:approx})
carry well-defined weak phases. The weak phase of $T$ is ${\rm
Arg}(V_{ud}V^*_{ub})=\gamma$, and that of $P+{2\over 3}P^C_{EW}$ is
approximately ${\rm Arg}(V_{td}V^*_{tb})=-\beta$, where we neglect corrections
due to quarks other than the top quark. The effects of the $u$ and $c$ quarks
become appreciable \cite{BF} when $V_{td}$ obtains its currently allowed
smallest values. This corresponds to a small deviation of the CP asymmetry in
$B^0(t)\to\pi^+\pi^-$ from $\sin(2\alpha)\sin(\Delta mt)$ (where $\Delta m$ is
the neutral $B$ mass-difference). For large values of $V_{td}$, where the
deviation due to the penguin amplitude becomes significant \cite{MG}, the $u$
and $c$ contributions become very small. $T'$ also carries the phase $\gamma$,
while the weak phase of $P'+{2\over 3}P'^C_{EW}$ is ${\rm
Arg}(V_{ts}V^*_{tb})=\pi$.

In what follows we shall denote $\cat \equiv |T|$, $\cap \equiv |P+{2\over
3}P^C_{EW}|$, $\catp \equiv |T'|$, $\capp \equiv |P'+{2\over 3}P'^C_{EW}|$. The
ratio of $\Delta S = 1$ to $\Delta S = 0$ tree and penguin amplitudes are given
by the corresponding ratios of CKM factors, $\catp/\cat =|V_{us}/V_{ud}| \equiv
r_u=0.23$, $\capp/\cap=|V_{ts}/V_{td}|\equiv r_t$. To introduce first-order
SU(3) breaking corrections, we note that in the $|T'|$ amplitude the $W$ turns
into an $\overline{s}$ quark instead of a $\overline{d}$ in $\cat$. Assuming
factorization for $T$, which is supported by experiments \cite{Browder,BS} and
justified for $B \to \pi \pi$ and $\pi K$ by the high momentum with which the
two color-singlet mesons separate from one another, SU(3) breaking is given by
the $K/\pi$ ratio of decay constants 
\beq
{{\cal T'}\over {\cal T}}={|V_{us}|\over |V_{ud}|}{f_K\over f_{\pi}}\equiv
\tilde{r}_u~~.
\eeq

Apart from small electroweak penguin terms, all amplitudes we consider are free
of color-suppressed contributions, for which factorization might be more
questionable.  The situation would be very different were we to consider the
amplitude for $B^0 \to \pi^0 \pi^0$, where the color-suppressed contribution
could be dominant.

In the penguin amplitudes (including electroweak penguin) of both $B^0\to\pi^-
K^+$ and $B^+\to\pi^+ K^0$ the $\overline{b}$ quark turns into an
$\overline{s}$ quark instead of a $\overline{d}$ in $B^0\to\pi^+\pi^-$. Here we
will denote the magnitude of the $\Delta S = 1$ penguin amplitude by $r_t
\tcap$, to allow for SU(3) breaking. Since factorization is questionable for
penguin amplitudes, one generally expects $\tcap \ne(f_K/f_{\pi}) \cap$. We
will assume that the phase $\delta_P$ is unaffected by SU(3) breaking. Since
this phase is likely to be small \cite{Penphase}, this assumption is not
expected to introduce a significant uncertaintly in the determination of the
weak phases. 

Assigning SU(3)-symmetric strong phases $\delta_T,~\delta_P$ to terms with
specific weak phases, and taking account of SU(3) breaking,
Eqs.~(\ref{eqn:amps}) and (\ref{eqn:approx}) may be transcribed as 

$$
\app ={\cal T}e^{i\delta_T}e^{i\gamma}+{\cal P}e^{i\delta_P}e^{-i\beta}~~,
$$
$$
\apk =\tilde{r}_u{\cal T}e^{i\delta_T}e^{i\gamma}-r_t\tilde{\cal P}
e^{i\delta_P}~~,
$$
\beq \label{eqn:withbk}
A_+=r_t\tilde{\cal P}e^{i\delta_P}~~.
\eeq

It will be shown that the numerous {\it a priori} unknown parameters in
(\ref{eqn:withbk}), including the two weak phases
$\alpha\equiv\pi-\beta-\gamma$ and $\gamma$, can be determined from the rate
measurements of the above three processes and their charge-conjugates. 

The amplitudes for the corresponding charge-conjugate decay processes are
simply obtained by changing the signs of the weak phases $\gamma$ and $\beta$.
We denote the charge-conjugate amplitudes corresponding to (\ref{eqn:withbk})
by $\overline{A}_{\pi\pi},~\overline{A}_{\pi K},~A_-$, respectively. A state
initially tagged as a $B^0$ or $\overline{B}^0$ will be called $B^0(t)$ or
$\overline{B}^0(t)$.  The time-dependent decay rates of these states to
$\pi^+\pi^-$ are given by 
$$
\Gamma(B^0(t)\to\pi^+\pi^-)=e^{-\Gamma t}[|\app|^2\cos^2({\Delta m\over
2}t)+|\abpp|^2\sin^2({\Delta m\over 2}t)
$$
$$+{\rm Im}(e^{2i\beta} \app \abpp^*)\sin(\Delta mt)]~~,
$$
$$
\Gamma(\overline{B}^0(t)\to\pi^+\pi^-) = e^{-\Gamma t}[|\app|^2\sin^2({\Delta
m\over 2}t) + |\abpp|^2\cos^2({\Delta m\over 2}t) 
$$
\beq
-{\rm Im}(e^{2i\beta} \app \abpp^* \sin(\Delta mt)]~~.
\eeq
Measurement of these quantities determines $|\app|^2,~|\abpp|^2$ and ${\rm
Im}(e^{2i\beta} \app \abpp^*)$.  It is convenient to define sums and
differences of the first two quantities, and we find 
$$
A \equiv \frac{1}{2}(|\app|^2 + |\abpp|^2)
= \cat^2 + \cap^2 - 2 \cat \cap \cos \delta \cos \alpha~~~,
$$
$$
B \equiv \frac{1}{2}(|\app|^2 - |\abpp|^2)
= - 2 \cat \cap \sin \delta \sin \alpha~~~,
$$
\beq \label{eqn:ABC}
C \equiv {\rm Im}~(e^{2 i \beta} \app \abpp^*) = - \cat^2 \sin 2\alpha
+ 2 \cat \cap \cos \delta \sin \alpha~~~,
\eeq
where we use $\beta + \gamma = \pi-\alpha$ and where we define $\delta \equiv
\delta_T - \delta_P$.

The rates of the self-tagging modes $\pi^- K^+,~\pi^+ K^-$ and $\pi^+K^0$ or
$\pi^- \overline{K}^0$ determine $|\apk|^2,~|\abpk|^2$ and $|A_+|^2$,
respectively.  Again, we take sums and differences of the first two, and find 
$$
D \equiv \frac{1}{2}(|\apk|^2 + |\abpk|^2) = (\tru \cat)^2 + \tcapp^2 - 2 \tru
\cat \tcapp \cos \delta \cos \gamma~~~~,
$$
$$
E \equiv \frac{1}{2}(|\apk|^2 - |\abpk|^2) = 2 \tru \cat \tcapp \sin \delta
\sin \gamma~~~,
$$
\beq \label{eqn:DEF}
F \equiv |A_+|^2 = |A_-|^2 = \tcapp^2~~~.
\eeq
The rates for $B^+ \to \pi^+ K^0$ and $B^- \to \pi^- \overline{K}^0$ are
expected to be equal, since only penguin amplitudes are expected to
contribute to these processes.  Here we have defined $\tcapp \equiv r_t
\tcap$.

Measurement of the six quantitities $A - F$ suffices to determine all six
parameters $\alpha,~\gamma,~\cat,~\cap,~ \tcap,~\delta$ up to discrete
ambiguities. The CKM parameter $r_t\equiv|V_{ts}/V_{td}|$, which is still
largely unknown, is obtained from the unitarity triangle in terms of $\alpha$
and $\gamma$: 
\beq \label{eqn:rurt}
r_u r_t={\sin\alpha\over\sin\gamma}~~.
\eeq

We note that
\beq \label{eqn:diffs}
|\apk|^2-|\abpk|^2=
-({f_K\over f_{\pi}})({\tilde{\cal P}\over{\cal P}})(|\app|^2 - |\abpp|^2)~~, 
\eeq
which determines the magnitude of SU(3) breaking in the penguin amplitude,
$\tcap/\cap$. The relation (\ref{eqn:diffs}) between the particle-antiparticle
rate differences in $B\to\pi K$ and in $B\to\pi\pi$ was recently derived
\cite{DH} in the SU(3) limit, $f_K/f_{\pi}\to 1,~\tilde{\cal P}/{\cal P}\to 1$.
The authors assumed for SU(3) breaking a value $\tilde{\cal P}/{\cal
P}=f_K/f_{\pi}$ (based on factorization of penguin amplitudes) which in our
approach is a free parameter to be determined by experiment. We expect it to
differ from one by up to 30$\%$. 

Both sides of Eq.~(\ref{eqn:diffs}) are proportional to $\sin \delta$, and thus
would vanish in the absence of a strong phase difference.  In that case, one
would have to assume a relation between $\tilde{\cal P}$ and ${\cal P}$ or some
other constraint in order to obtain a solution.  If, on the other hand, $\delta
\ne 0$, leading to a rate asymmetry between the self-tagging decays $B^0 \to
\pi^- K^+$ and $\overline{B}^0 \to \pi^+ K^-$, the present method permits one
to interpret that rate asymmetry in a manner independent of $\delta$. 
\bigskip

\leftline{\bf B.  Likely ranges of observables}
\bigskip

The amplitude for $B^0 \to \pi^+ \pi^-$ is expected to be dominated by the
$\cat$ contribution, while those for $B^0 \to \pi^- K^+$ and $B^+ \to
\pi^+ K^0$ are expected to be dominated by $\tcapp$. We shall choose units in
which a branching ratio of $10^{-5}$ corresponds to a value of 1 for the rates
$A$, $D$, and $F$.  The normalizations of the other quantities are set
accordingly.  We shall also define the quantity 
\beq \label{eqn:Sdef}
S \equiv A + D = \frac{1}{2}(|\app|^2 + |\abpp|^2 + |\apk|^2 + |\abpk|^2)
~~~.
\eeq

A combined sample of the decays $B^0\to\pi^+\pi^-$ and $B^0\to\pi^- K^+$ has
been observed with a joint branching ratio of $(1.8^{+0.6+0.2}_{-0.5-0.3}\pm
0.2) \times 10^{-5}$ \cite{piK}, so that $S = 1.8 \pm 0.65$. Equal mixtures of
the two modes are most likely, corresponding to individual branching ratios of
about $10^{-5}$ for $B^0\to\pi^+\pi^-$ and $B^0\to\pi^- K^+$. A similar
branching ratio is expected for $B^+\to\pi^+ K^0$ if the $\tcapp$ amplitude
dominates $B \to \pi K$ decays, as seems likely. Thus, values of order 1 for
$A$, $D$, and $F$ are expected.  We shall consider a range of values for these
quantities, subject only to constraints on the lower and upper limits for $S$.
As we shall see in Sec.~III D, when $\delta$ can be neglected, this works out
to a rule of thumb that $\cat^2 + \tcapp^2 \approx 2$. 

The detection of $B^+ \to \pi^+ K^0$ or $B^- \to \pi^- \bar K^0$ in practice
will utilize the channels $B^\pm \to \pi^\pm K_S,~K_S \to \pi^+ \pi^-$, with a
corresponding loss in efficiency of a factor of 3.  We shall take this factor
into account in estimating statistical errors on $F$. 

The remaining quantities $B$, $C$, and $E$ are harder to anticipate.
The Schwarz inequality limits the value of $|C|$ to be less than or equal
to $A$.  In practice we find values of $|C|$ larger than 2 to be very
unlikely.  Thus, we shall consider values subject to this restriction.
Both $B$ and $E$ will vanish if $\delta = 0$.  While a recent calculation
\cite{KP} based on perturbative QCD \cite{Penphase} suggests that $\delta_T
\approx 0$, $\delta_P \approx 9.5^{\circ}$, $\delta \approx - 9.5^{\circ}$, 
the possibility of non-perturbative effects (such as strong final-state
interactions differing in channels of different isospin) cannot be excluded.
Thus, we shall consider the representative values $\delta =
0,~5.7^\circ,~36.9^\circ,~ 84.3^\circ,~95.7^\circ,~143.1^\circ,~174.3^\circ$. 
We take only non-negative values since the error estimates are not affected by
sign changes in $\delta$. The nonzero values will be discussed in Sec.~IV. 
\bigskip

\leftline{\bf C.  Constraints on the angles $\alpha$ and $\gamma$}
\bigskip

Recent analyses of constraints on the CKM parameters include those in
Refs.~\cite{AL} and \cite{values}.  We shall visualize the allowed ranges of
$\alpha$ and $\gamma$ in order to choose illustrative sets of parameters. 

We begin with the Wolfenstein parameterization \cite{WP} of CKM elements:
\beq \label{eqn:WP}
V_{cb} = A \lambda^2~~~,~~~
V_{ub} = A \lambda^3 (\rho - i \eta)~~~,~~~
V_{td} = A \lambda^3 (1 - \rho - i \eta)~~~,
\eeq
as well as others not quoted explicitly, where $\lambda = 0.22$. We shall
assume \cite{values} $V_{cb} = 0.038 \pm 0.003$.  (A slightly higher value is
quoted in Ref.~\cite{PDG}.)  The measurement \cite{values} $|V_{ub}/V_{cb}| =
0.08 \pm 0.02$ based on charmless $B$ decays implies $(\rho^2 + \eta^2)^{1/2} =
0.36 \pm 0.09$.  The measurement of $B^0 - \overline{B}^0$ mixing implies
\cite{PDG} $|V_{td}| = 0.009 \pm 0.003$, which we shall interpret as implying
$|1- \rho - i \eta| = 1.0 \pm 0.3$.  The requirement that the imaginary part of
the $K^0 - \overline{K}^0$ mixing amplitude due to CKM phases be responsible
for the observed CP violation in the kaon system implies a hyperbola
\cite{values} $\eta(1 - \rho + 0.35) = 0.48 \pm 0.20$, where the $1 - \rho$
term in parentheses refers to the contribution of the top quark loop, while
0.35 refers to the charmed quark's contribution. 

The allowed region in $(\rho,\eta)$ is shown in Fig.~1(a); the corresponding
range of $(\alpha,\gamma)$ is depicted in Fig.~1(b) by a rather narrow band. 
The strong anticorrelation between $\alpha$ and $\gamma$ is a function of the
limited range of $\beta = \pi - \alpha - \gamma$, which is restricted to
$6.6^\circ \le \beta \le 27^\circ$ for the present set of parameters
\cite{values}.  The band would not be quite so narrow using the parameters of
one other analysis \cite{AL}.


Three representative points, noted on the figure, are described in Table I. 
These correspond to extreme and central values of $\alpha$ and $\gamma$.  A
number of illustrative examples will be presented for these points. 

Also shown on Fig.~1(b) is a linear least-squares fit to the points $p_1$,
$p_2$, and $p_3$: $\gamma = 175^{\circ} - 1.16 \alpha$. Almost equally good is
the approximate relation 
\beq \label{eqn:line}
\gamma = 180^{\circ} - 1.2 \alpha~~~,
\eeq
which we shall use in Sec.~III A to simplify relations between $\pi \pi$ and
$\pi K$ rates.

\begin{table}
\caption{Representative points in the $(\rho,\eta)$ plane and corresponding
angles of the unitarity triangle.} 
\begin{center}
\begin{tabular}{c c c c c c c} \hline
Point & $\rho$ & $\eta$ & $\alpha$ & $\beta$ & $\gamma$ &  $r_t$  \\
      &        &        &  (deg.)  &  (deg.) &  (deg.)  &         \\ \hline
$p_1$ &$-0.30$ &  0.15  &   20.0   &   6.6   &  153.3   &   3.36  \\ 
$p_2$ &   0    &  0.35  &   70.7   &  19.3   &   90.0   &   4.16  \\
$p_3$ &  0.36  &  0.27  &  120.3   &  22.9   &   36.9   &   6.35  \\ \hline
\end{tabular}
\end{center}
\end{table}
\bigskip

\leftline{\bf D.  Limitation associated with size of $\delta$}
\bigskip

The relation (\ref{eqn:diffs}) can be written as
\beq
\frac{E}{B} = - \frac{f_K}{f_\pi}~\frac{\tcapp}{\capp}~~~.
\eeq

We wish to evaluate the ratio $\tcapp/\capp$ to better than 30\% (the
anticipated magnitude of SU(3) breaking).  In this subsection we estimate the
number of $\pi^+ \pi^-$ events (and, correspondingly, the size of the other
data samples) needed in order to measure $B$ and $E$ [see the expressions
(\ref{eqn:ABC}) and (\ref{eqn:DEF})] to the required accuracy. 

Define the number of events averaged between particle and antiparticle decays:
$$
\frac{N(B^0 \to \pi^+ \pi^-) + N(\overline{B}^0 \to \pi^+ \pi^-)}{2} \equiv
N_{\pi \pi}~~~,
$$
\beq
\frac{N(B^0 \to \pi^- K^+) + N(\overline{B}^0 \to \pi^+ K^-)}
{2} \equiv N_{\pi K}~~~. 
\eeq 
With equal branching ratios for $\pi^+ \pi^-$ and $\pi^\pm K^\mp$, the present
data sample would consist of about 10 events each for $N_{\pi \pi}$ and $N_{\pi
K}$ \cite{piK}.  The errors on $A$ and $B$ both scale as $N_{\pi \pi}^{1/2}$,
while those on $D$ and $E$ scale as $N_{\pi K}^{1/2}$.  Then in the samples of
events used to measure $A$, $B$, $D$, and $E$, we expect 
\beq
\delta N_A \simeq \delta N_B \simeq N_{\pi \pi}^{1/2}~~,~~
\delta N_D \simeq \delta N_E \simeq N_{\pi K}^{1/2}~~~.
\eeq

We take as illustrative parameters $\cat = \tcapp = \capp = 1$, neglecting
SU(3) breaking in the ratio $\tcapp/\capp$. Recalling the expressions for $B$
and $E$, we expect the numbers of events in the samples corresponding to these
quantities to be 
\beq
N_B = - 2 N_{\pi \pi} r_u \sin \delta \sin \gamma~~~,~~~
N_E = 2 N_{\pi K} \tru \sin \delta \sin \gamma~~~,
\eeq
where we have used (\ref{eqn:rurt}).  Consequently, the fractional errors on
$N_B$ and $N_E$ are 
\beq
\frac{|\delta N_B|}{|N_B|} \simeq \frac{1}{2 N_{\pi \pi}^{1/2} r_u |\sin \delta
\sin \gamma|}~~~,
\eeq
\beq
\frac{|\delta N_E|}{|N_E|} \simeq \frac{1}{2 N_{\pi \pi}^{1/2}\tru |\sin \delta
\sin \gamma|}~~~.
\eeq
Thus the fractional error on the quotient $N_B/N_E$ is the sum in quadrature of
these two errors:
\beq
\frac{|\delta(N_B/N_E)|}{|(N_B/N_E)|} = \frac{1}{2 N_{\pi
\pi}^{1/2} r_u |\sin \delta \sin \gamma|} \left( 1 + \frac{f_\pi^2}{f_K^2}
\right)^{1/2}~~~.
\eeq
Demanding that this error be less than 30\% as noted above and substituting the
values of the constants, we find 
\beq \label{eqn:statest}
N_{\pi \pi} \ge \frac{91}{\sin^2 \delta \sin^2 \gamma}~~~.
\eeq
This gives an idea of the data samples required to improve upon the assumption
of no more than 30\% SU(3) breaking in penguin amplitudes. More detailed
estimates are postponed until Sec.~IV. Meanwhile, we examine the special case
in which the strong phase shift difference $\delta$ vanishes. 
\bigskip

\centerline{\bf III.  VANISHING STRONG PHASE SHIFT DIFFERENCE}
\bigskip

Both the $\pi \pi$ parameter $B$ and the $\pi K$ rate asymmetry parameter $E$
vanish when the strong phase shift difference $\delta$ is zero. In that case,
however, one can no longer use the relation (\ref{eqn:diffs}) to determine the
ratio $\tcap/\cap$.  One has 4 observables ($A$, $C$, $D$, and $F$) to
determine 5 parameters (e.g., $\cat$, $\cap$, $\alpha$, $\tcapp$, and
$\gamma$).  One must make additional assumptions to obtain solutions.  In this
section we explore several such possibilities. 
\bigskip

\leftline{\bf A.  Simplified observables with $\delta = 0$.}
\bigskip

When $\delta = 0$, the equations in (\ref{eqn:ABC}) and
(\ref{eqn:DEF}) for $A,~C,$ and $D$ become 
$$
A = \cat^2 + \cap^2 - 2 \cat \cap \cos \alpha~~~,
$$
$$
C = - \cat^2 \sin (2\alpha) + 2 \cat \cap \sin \alpha~~~,
$$
 \beq \label{eqn:ABD}
D = (\tru \cat)^2 + \tcapp^2 - 2 \tru \cat \tcapp \cos \gamma~~~.
\eeq

A simple relation follows from eliminating $\cap$ between the first two of
these equations: 
\beq
(C/A)^2 = 4 z (1 - z)~~~,~~~{\rm where}~~~z \equiv \cat^2 \sin^2 \alpha/A~~~.
\eeq
The Schwarz inequality bound $|C/A| \le 1$ mentioned earlier is manifest here. 
\bigskip

\leftline{\bf B.  Linear relation between $\gamma$ and $\alpha$}
\bigskip

A considerable simplification useful for anticipating the precision in
determining $\alpha$ and $\gamma$ is obtained by noting that $\alpha$ and
$\gamma$ are rather tightly correlated with one another [see Fig.~1(b)] as a
result of the restricted range of $\beta$.  As mentioned, the dependence can be
approximated by a straight line.  If we are concerned mainly with learning the
sign of $\rho$ and are not so concerned about the exact magnitude of $(\rho^2 +
\eta^2)^{1/2}$, we can substitute for $\gamma$ in the expression
(\ref{eqn:ABD}) for $D$, having already substituted $F = \capp^2$, and thus
each measurement of $D$ implies a relation between $\cat$ and $\alpha$. 

An even greater simplification can be obtained if we neglect the term
quadratic in $\cat$ in $D$, and eliminate $\cat$, $\cap$, and $\tcapp$ from the
remaining equations involving $A$, $C$, $D$, and $F$.  With the approximate
formula (\ref{eqn:line}) one has
\beq \label{eqn:oneangle}
\frac{\cos(1.2 \alpha)}{\sin \alpha} = 2.6 \frac{D-F}{\sqrt{F}}\frac{1}
{\sqrt{A \pm \sqrt{A^2 - C^2}}}~~~.
\eeq
The sign ambiguity stems from the fact that the equation for $\cat \sin \alpha$
has two solutions.  We can anticipate that the solution with $\cat^2 \approx A$
is the most likely, as long as $\cap$ is relatively small compared to $\cat$,
as generally anticipated.  Since 
\beq
(\cat \sin \alpha)^2 = \frac{A \pm \sqrt{A^2 - C^2}}{2}~~~,
\eeq
and since $|C|$ tends to be small when $\sin \alpha$ is near 1 (as for the
point $p_2$), we anticipate that in that case we should choose the positive
sign in the square root, and the argument of the overall square root in the
denominator of the last fraction in Eq.~(\ref{eqn:oneangle}) is about $2A$.  On
the other hand, when $|C|$ is fairly large (e.g., for points $p_1$ and $p_3$),
the sign does not matter much, and the argument of the overall square root is
about $A$. 

For 100 $\pi^+ \pi^-$ and 100 $\pi^\pm K^\mp$ events, the errors in $A$ and $D$
are about 10\%.  Assuming that the $|P'|$ contribution is dominant in $D$, the
error on $F$ will then be about 17\% (because of the branching ratio of neutral
kaons to $\pi^+ \pi^-$).  One then finds an error of about 0.55 in the
right-hand side of Eq.~(\ref{eqn:oneangle}) when $\alpha$ is near the middle of
its range, and about 0.78 when $\alpha$ is near its lower or upper bounds. In
Fig.~2(a) we plot the left-hand side of this equation, along with plotted
points for $\alpha = 20^{\circ}~(p_1),$ $71^{\circ}~(p_2),$ and
$120^{\circ}~(p_3)$, with the errors in $\cos(1.2 \alpha)/ \sin \alpha$ of $\pm
(0.78,~0.55,~0.78)$, respectively.  The allowed region in $(\alpha,\gamma)$ is
shown in Fig.~2(b) along with the line corresponding to $\gamma = 180^{\circ} -
1.2 \alpha$.  The arrows designate values of $\alpha$ corresponding to $p_1$,
$p_2$, and $p_3$. 

When $\alpha$ is close to the center of its range, a sample of $B$ decays
corresponding to 100 events in each of the $\pi^+ \pi^-$ and $\pi^\pm K^\mp$
channels allows one to narrow the allowed region of $\alpha$ by roughly a
factor of 2.  For the lowest or highest allowed values of $\alpha$ one does
somewhat better.  For more precise estimates, one would retain the $|T|^2$ term
in $D$ when $\alpha \simeq 90^{\circ}$, and would be more precise about the
error on $C$.

Let us for the moment neglect the small correction term in the expression
(\ref{eqn:ABC}) for $C$.  Then since $\sin 2 \alpha$ can take on the same value
for two values of $\alpha$ equally above and below $\pi/4$, there is a discrete
ambiguity associated with negative values of $C$ and values of $\alpha <
\pi/2$.  This ambiguity is likely to persist when we include the correction
term.  However, the addition of $\pi K$ decay information appears capable of
resolving this ambiguity, since it provides additional information on the angle
$\gamma$ which is highly correlated with $\alpha$. For values of $\alpha >
\pi/2$ and positive $C$ only one solution (that for $\alpha < 3 \pi/4$) appears
to be in the physical region, so we do not get the same sort of discrete
ambiguity. 
\bigskip

\leftline{\bf C.  SU(3) assumption for penguin amplitudes}
\bigskip

When $\delta=0$, as mentioned, we are missing information on $\tcapp/\capp$. In
the previous subsection we supplied this information by assuming a functional
relation between $\alpha$ and $\gamma$.  In the present subsection, we no
longer assume such a relation, but simply assume this ratio to equal unity. 
(It was assumed to equal $f_K/f_\pi \approx 1.2$ in Ref.~\cite{SilWo} and
\cite{DH}.) Under this assumption, we may drop the tilde symbols on $\tcapp$ in
Eqs.~(\ref{eqn:ABD}).  We may then plot contours of observables in the plane of
$\cat \equiv |T|$ vs. $\capp \equiv |P'|$ for various regions in the allowed
parameter space. 
 
Contours of fixed $A$ are mainly sensitive to $\cat$, while those of fixed $D$
depend mainly on $\capp$.  The slopes of the contours reflect the presence of
constructive or destructive interference between $\cat$ and $\capp$, depending
on the signs of $\cos \alpha$ and $\cos \gamma$.  Because of the strong
anticorrelation between $\alpha$ and $\gamma$ shown in Fig.~1, the contours of
$A$ and $D$ are nearly perpendicular to one another for each of the three
illustrated cases.  This means that for each pair $(\alpha,\gamma)$, a
measurement of $A$ and $D$ selects a point in the $(\cat,\capp)$ plane with
comparable errors on $\cat$ and $\capp$ if the values and errors of $A$ and $D$
are comparable to one another. 

A measurement of $F = |\capp|^2$ (the $\pi^+ K^0$ or $\pi^- \overline{K}^0$
branching ratio, in units of $10^{-5}$) must be consistent with the
determination just made.  Thus, a one-dimensional allowed set of points
$(\alpha,\gamma)$ is chosen by the combined measurements of $A$, $D$, and $F$. 
The degree to which this choice is unique depends on being able to observe the
effect of $\cat - \capp$ interference in the measurement of $D$, since in the
absence of the contribution from $\cat$ one would have $D = F$.  That is, the
average $\pi^\pm K^\mp$ and $\pi^+ K^0$ (or $\pi^- \overline{K}^0$) rates would
be the same in the absence of the $\cat$ contribution to the $\pi^\pm K^\mp$
mode. 

Once one has selected values of $\cat$ and $\capp$ for a one-dimensional set of
points in the $(\alpha,\gamma)$ plane, the value of $C$ can be used to
distinguish among those points. Positive values of $C$ tend to be associated
with negative values of $\sin (2 \alpha)$ and hence with values of $\alpha$
greater than $90^\circ$. Such values correspond to $(\rho,\eta)$ values lying
inside a circle of radius 1/2 with center at $\rho = 1/2,~\eta = 0$.  These
parameters correspond to roughly the right-hand one-third of the allowed
regions in Figs.~1(a) and 1(b).  The parameter spaces of Fig.~1 are much more
sensitive to positive values of $C$ than to negative values.  We shall see such
behavior again when we come to discuss a further simplification in the next
subsection. 
\bigskip

\leftline{\bf D.  Information without $\pi/K$ separation}
\bigskip

The fact that the contours of $A$ and $D$ are nearly perpendicular to one
another and have similar spacings in $\cat$ and $\capp$, respectively, suggests
that contours of 
\beq
S = A + D = (1+\tilde{r}^2_u)\cat^2 + (1+r^{-2}_t)\capp^2 - 2\cat \capp
(r^{-1}_t\cos\alpha + \tilde{r}_u\cos\gamma)
\eeq
may not depend very much on which set of allowed $(\alpha,\gamma)$ one chooses.

This expectation is borne out in Fig.~3. The observation that $S$ is roughly
independent of $(\alpha,\gamma)$ follows from the anticorrelation of $\cos
\alpha$ and $\cos \gamma$. The sum of the average $B^0$ and $\overline{B}^0$
branching ratios to $\pi^+ \pi^-$ and $\pi^\pm K^\mp$ leads to an approximate
constraint on $\cat^2 + \capp^2$ roughly independent of CKM parameters within
the allowed range.  This is fortunate, since the CLEO Collaboration \cite{piK}
measures precisely this sum, with only weak distinction at present between
pions and kaons. (Improved particle identification at CLEO is foreseen in the
future.)  One may expect a similar measurement in some hadron collider
experiments, such as the CDF Detector at Fermilab, unless specific steps are
taken for particle identification. 

A combined measurement of $\cat^2 + \capp^2$ and $F = \capp^2$ now can be used
to determine each parameter.  The determination of $\alpha$ and $\gamma$ is now
more simple-minded, albeit less precise, than in the previous section.  Since
our determination of $\cat$ and $\capp$ is now independent of $\alpha$ and
$\gamma$, we can use these parameters in the equation (\ref{eqn:ABD}) for $C$
to plot contours of fixed $C$ in the $(\alpha, \gamma)$ plane.  (The variation
with $\gamma$ springs from the fact that $r_u r_t = \sin \alpha / \sin \gamma$,
as mentioned previously.)  An example of such a contour is plotted in Fig.~4
for the representative values $\cat = \capp = 1.1$.  One sees a fair amount of
uncertainty for values of $C$ near $-0.7$, where the contour intersects the
allowed $(\alpha,\gamma)$ region in a wide range of points.  This behavior is
related to the discrete ambiguity noted at the end of Sec.~III B.  However,
contours of positive $C$ cut the allowed region at a larger angle and lead to a
more highly constrained solution. 

The measurement of $C$ in the absence of particle identification is possible
since one is following the time-dependence of a decay rate in which one
compares the decays of states which were initially $B^0$ and $\overline{B}^0$
to a combination of final states.  The oscillations from which $C$ is to be
extracted are expected to stem only from the $\pi^+ \pi^-$ final state. 

The imposition of particle identification returns one to the situation of the
previous section.  As mentioned, to make efficient use of the information thus
provided, one must be able to see the difference between $D$, where there is a
small $\cat$ contribution, and $F$, where there is none. 
\bigskip

\leftline{\bf E.  Results of a Monte Carlo simulation}
\bigskip

We have explored numerically the case of $\delta = 0,~\tcapp = \capp$ using a
Monte Carlo program which generates events with a statistical spread in the
variables $A,~C,~D$, and $F$ appropriate to data samples corresponding to a
total of $M$ decays of $B^0$ or $\overline{B}^0$ to $\pi^+ \pi^-$. Scaling
other quantities to the expected $\pi^+ \pi^-$ rates, and recalling the
discussion of Sec.~II D, we then assume 
\beq \label{eqn:errs}
\delta A = \delta C = \delta D = \delta F/\st = 1/\sqrt{M}~~~.
\eeq
For $M = 100,~1000,~10000$ we then ask how well $\alpha$ and $\gamma$ can
be determined.  Numerically this is accomplished by stepping $\alpha$ and
$\gamma$ through a range of values, accepting any solution which is within
$2 \sigma$ of the generated value for each parameter $A,~C,~D,~F$, and
averaging all such solutions.  To allow for the possibility of multiple
solutions, a cluster algorithm (described in the Appendix) is applied.
We also restrict $10^\circ \le \alpha \le 130^\circ$, $20^\circ \le \gamma
\le 170^\circ$, and $|\gamma - (175^\circ - 1.16 \alpha)| \le 30^\circ$ in
accord with the allowed regions in Fig.~1.  The results are shown in Fig.~5.

One sees a noticeable improvement with increased statistics.  A clear
distinction between the cases ($p_1$ or $p_2$) and $p_3$ is already possible
with $M = 100$.  Satisfactory results for all three points are obtained for $M
= 1000$. The small ambiguity for $p_1$ appears to be related to the multiple
intersections of the contours for negative $C$ (Fig.~4) with the allowed region
of parameter space.  Errors are reduced further when $M$ is increased to 10000. 
\newpage

\centerline{\bf IV.  ERROR ANALYSIS FOR GENERAL FINAL-STATE PHASES}
\bigskip

In Ref.~\cite{GRPRL} we estimated the statistical accuracy of determining the
weak phases $\alpha$ and $\gamma$ using the present method to be at a level of
ten percent, given around a hundred events in each channel.  The theoretical
uncertainty of the method is at a similar level, involving the following
corrections all of which are of order a few percent:  A correction from an
electroweak penguin amplitude in $B^+\to\pi^+ K^0$, corrections due to $u$ and
$c$ quarks in the $B^0\to\pi^+\pi^-$ penguin amplitude, second-order SU(3)
breaking in the magnitudes of weak amplitudes, first order SU(3) breaking in
the (small) strong phase of the penguin amplitude, and ${\cal O}(f_B/m_B)$
annihilation amplitudes.  In this subsection we analyze in more detail the
precision on $\alpha$ and $\gamma$ that can be attained with a given sample of
events as a function of the parameters.  We use a Monte Carlo program similar
to that described in Sec.~III E to generate events with a given Gaussian
distribution in the parameters $A - F$ appropriate to a total sample
corresponding to $M~\pi^+ \pi^-$ decays.  We choose $\cat = 1$ and $\tcapp =
\capp = 1$ for the purpose of generating events.  In addition to the errors
assumed in Eq.~(\ref{eqn:errs}), we assume 
\beq \label{eqn:beers}
\delta B = \delta E = 1/\sqrt{M}~~~.
\eeq
The results are shown in Fig.~6.

The method clearly improves with increased statistics, in a manner roughly
compatible with our estimate (\ref{eqn:statest}).  Despite the presence of
a large spread in values of $\alpha$ and $\gamma$, one can already distinguish
the case $p_3$ from the cases $p_1$ and $p_2$ for $M = 100$.  The distinction
between $p_1$ and $p_2$ appears to emerge by the time one reaches $M = 1000$.
On the other hand, one sees the distinct appearance of clusters of points,
corresponding to the presence of discrete ambiguities.  These ambiguities
are to be distinguished from the one mentioned at the end of Sec.~III B, and
appear to be related to uncertainties in the value of the final-state phase.
Their nature is best ascertained by referring to the case $M = 10000$. One can
detect up to three clusters of solutions.  For example, in the case of input
parameters in the vicinity of $p_1$, a final-state phase $\delta \approx
90^\circ$ turns out to be ambiguous with two other cases, one with $\delta <
90^\circ$ and the other with $\delta > 90^\circ$, as one sees by referring
to the corresponding plots of $\alpha$ vs. $\delta$ shown in Fig.~7. For input
parameters near $p_2$, the more serious ambiguities appear to occur for
moderate or small values of $\delta$.  For input parameters near $p_3$,
uncertainties present for $M = 100$ and $M = 1000$ appear to have largely
disappeared by the time $M$ reaches 10000.  This behavior may be related to the
uniqueness of the solution provided by large positive $C$ for points near $p_3$
in Fig.~4, but also points to the absence of ambiguities associated with the
value of $\delta$. 

The discrete ambiguities mentioned previously are quite noticeable in Fig.~7;
in addition, for $\delta$ near $180^\circ$, even the largest value of $M$ does
not lead to solutions in which that value is uniquely determined. 
\newpage

\centerline{\bf V.  CONCLUSIONS}
\bigskip

To summarize, we have shown that measurements of the rates for $B$ decays to
modes involving charged pions and kaons in the final states can determine the
shape of the unitarity triangle, even in the absence of theoretical or
experimental information about final-state phases.  The full set of
measurements involves the detection of the time-dependent rates for $B^0$ and
$\overline{B}^0 \to \pi^+ \pi^-$, and the rates for $B^0 \to \pi^- K^+$,
$\overline{B}^0 \to \pi^+ K^-$, and $B^\pm \to K_S \pi^\pm$.  A rate asymmetry
between $B^0 \to \pi^- K^+$ and $\overline{B}^0 \to \pi^+ K^-$ is needed in
order to perform a solution for all necessary parameters.  In the absence of
this asymmetry, one can obtain partial information by noting the tight
correlation between the angles $\alpha$ and $\gamma$ of the unitarity
triangle, or by assuming an SU(3) relation between strangeness-changing and
strangeness-preserving penguin amplitudes.  One can even dispense with
particle identification, summing $\pi^+ \pi^-$ and $\pi^\pm K^\mp$ modes, if
only crude constraints on parameters are desired.  As a result of the strong
anticorrelation between $\alpha$ and $\gamma$ in the physically allowed region
of parameter space, the $\pi K$ modes are particularly helpful in resolving a
discrete ambiguity associated with the behavior of the function $\sin 2 \alpha$
which would be present if one studied $\pi \pi$ modes alone. 

In the simplest case examined, where the assumption of $\delta = 0$ and the
strong correlation between $\alpha$ and $\gamma$ in the allowed parameter space
were utilized, we found that a sample of events corresponding to 100 $\pi^+
\pi^-$ and 100 $\pi^\pm K^\mp$ events, with a correspondingly reduced number of
detected $B^\pm \to K_S \pi^\pm$ decays, was sufficient to reduce the allowed
region in parameter space by roughly a factor of two, depending on the values
of the CKM angles. 

In the more general case in which $\delta \approx 0$ but no relation between
$\alpha$ and $\gamma$ was assumed, we found that by assuming SU(3) symmetry
for penguin amplitudes we could obtain unique solutions for $\alpha$ and
$\gamma$, with some possibility of discrete ambiguity when $\alpha$ is
small and $\gamma$ is large (corresponding to $\rho <0$ in the language
of the Wolfenstein parametrization).  Even when a distinction
between charged pions and charged kaons is not possible (Sec.~III D), partial
information on the parameters can be obtained, since the time-dependent
effects are expected to be confined to the $\pi^+ \pi^-$ channel and thus a
measurement of the parameter $C$ (defined in Sec.~II) is still possible.

In the most general case of nonzero final-state phase differences $\delta$ we
find that the program described here requires approximately $10^2/(\sin^2
\delta \sin^2 \gamma)$ decays of neutral $B$'s to charged pions (and a similar
number of $\pi K$ events) in order to free oneself from assumptions of SU(3)
breaking at the 30\% level in penguin amplitudes.  A Monte Carlo program has
shown that one begins to get useful information with 100 such decays (to be
compared with about 10 in the present data sample). The full power of the
method becomes apparent as the sample exceeds 1000 and approaches 10000.  Even
under such circumstances, a discrete ambiguity remains associated with the size
of final-state phases.  Arguments external to those presented here (such as the
allowed regions in the $(\alpha,\gamma)$ parameter space, the expected
magnitude of SU(3) breaking, and the expected size of final-state phases) may
be necessary to resolve such ambiguities.
\newpage

\centerline{\bf ACKNOWLEDGMENTS}
\bigskip

We thank J. Bjorken, F. De Jongh, H. Lipkin, D. London, H. Quinn, P. Sphicas,
and S. Stone for fruitful discussions, and the CERN Theory Group for a
congenial atmosphere in which part of this collaboration was carried out. A. D.
wishes to thank G. Harris for valuable advice on Monte Carlo methods.  M. G.
and J. L. R. wish to acknowledge the respective hospitalities of the SLAC and
Fermilab theory groups during parts of this investigation, and J. L. R. thanks
the Physics Department of the Technion for its hospitality.  This work was
supported in part by the United States -- Israel Binational Science Foundation
under Research Grant Agreement 94-00253/1, by the Fund for Promotion of
Research at the Technion, and by the United States Department of Energy under
Contract No. DE FG02 90ER40560. 
\bigskip

\centerline{\bf APPENDIX:  DETAILS OF CLUSTER ALGORITHM}
\bigskip

Given a set of values for $A, B, C, D, E$ and $F$, the values of $\alpha,
\gamma$ and $\delta$ are not necessarily determined uniquely. Apart from the
ambiguities associated with the numerical nature of the algorithm, there can
also be discrete ambiguities. In that case the set of triplets $(\alpha,
\gamma, \delta)$ consistent with all the observed quantities will form clusters
in the $(\alpha, \gamma, \delta)$ space, for any given set of $A, B, C, D, E,
F$. The number of clusters correspond to the number of discrete solutions and
the spread within a cluster corresponds to the (numerical) error on that
particular point. The average of all points in each cluster is taken to be the
{\it central value} for that cluster and is plotted in Figs.~5, 6 and 7 as a
single point. The number of points plotted for each data set is thus the number
of discrete solutions for that data set. 

The ambiguities associated with the numerical nature are expected to be {\it
continuous}; i.e. for any point $i$ to belong to a cluster, there should be at
least one point $j$ in the cluster such that $|\alpha_i - \alpha_j| \leq \Delta
\alpha$, where $\Delta \alpha$ is the {\it least count} in $\alpha$ in the
numerical algorithm. It is observed that this condition alone  is sufficient to
separate the clusters and additional conditions on $\gamma_i$ or $\delta_i$ are
not needed. Two points $i$ and $j$ belonging to different clusters will fail to
satisfy this condition. Different clusters can thus be separated from each
other.
\bigskip

\def \ajp#1#2#3{Am.~J.~Phys.~{\bf#1}, #2 (#3)}
\def \apny#1#2#3{Ann.~Phys.~(N.Y.) {\bf#1} #2 (#3)}
\def \app#1#2#3{Acta Phys.~Polonica {\bf#1} #2 (#3)}
\def \arnps#1#2#3{Ann.~Rev.~Nucl.~Part.~Sci.~{\bf#1} #2 (#3)}
\def \baps#1#2#3{Bull.~Am.~Phys.~Soc. {\bf#1} #2 (#3)}
\def \cmp#1#2#3{Commun.~Math.~Phys.~{\bf#1} #2 (#3)}
\def \cmts#1#2#3{Comments on Nucl.~Part.~Phys.~{\bf#1} #2 (#3)}
\def \cn{Collaboration}
\def \corn93{{\it Lepton and Photon Interactions:  XVI International Symposium,
Ithaca, NY August 1993}, AIP Conference Proceedings No.~302, ed.~by P. Drell
and D. Rubin (AIP, New York, 1994)}
\def \cp89{{\it CP Violation,} edited by C. Jarlskog (World Scientific,
Singapore, 1989)}
\def \dpff{{\it The Fermilab Meeting -- DPF 92} (7th Meeting of the American
Physical Society Division of Particles and Fields), 10--14 November 1992,
ed. by C. H. Albright \ite~(World Scientific, Singapore, 1993)}
\def \dpf94{DPF 94 Meeting, Albuquerque, NM, Aug.~2--6, 1994}
\def \efi{Enrico Fermi Institute Report No. EFI}
\def \el#1#2#3{Europhys.~Lett.~{\bf#1} #2 (#3)}
\def \f79{{\it Proceedings of the 1979 International Symposium on Lepton and
Photon Interactions at High Energies,} Fermilab, August 23-29, 1979, ed.~by
T. B. W. Kirk and H. D. I. Abarbanel (Fermi National Accelerator Laboratory,
Batavia, IL, 1979}
\def \hb87{{\it Proceeding of the 1987 International Symposium on Lepton and
Photon Interactions at High Energies,} Hamburg, 1987, ed.~by W. Bartel
and R. R\"uckl (Nucl. Phys. B, Proc. Suppl., vol. 3) (North-Holland,
Amsterdam, 1988)}
\def \ib{{\it ibid.}~}
\def \ibj#1#2#3{~{\bf#1}, #2 (#3)}
\def \ichep72{{\it Proceedings of the XVI International Conference on High
Energy Physics}, Chicago and Batavia, Illinois, Sept. 6--13, 1972,
edited by J. D. Jackson, A. Roberts, and R. Donaldson (Fermilab, Batavia,
IL, 1972)}
\def \ijmpa#1#2#3{Int.~J.~Mod.~Phys.~A {\bf#1} #2 (#3)}
\def \ite{{\it et al.}}
\def \jmp#1#2#3{J.~Math.~Phys.~{\bf#1} #2 (#3)}
\def \jpg#1#2#3{J.~Phys.~G {\bf#1} #2 (#3)}
\def \lkl87{{\it Selected Topics in Electroweak Interactions} (Proceedings of
the Second Lake Louise Institute on New Frontiers in Particle Physics, 15--21
February, 1987), edited by J. M. Cameron \ite~(World Scientific, Singapore,
1987)}
\def \ky85{{\it Proceedings of the International Symposium on Lepton and
Photon Interactions at High Energy,} Kyoto, Aug.~19-24, 1985, edited by M.
Konuma and K. Takahashi (Kyoto Univ., Kyoto, 1985)}
\def \mpla#1#2#3{Mod.~Phys.~Lett.~A {\bf#1} #2 (#3)}
\def \nc#1#2#3{Nuovo Cim.~{\bf#1} #2 (#3)}
\def \np#1#2#3{Nucl.~Phys.~B{\bf#1}, #2 (#3)}
\def \pisma#1#2#3#4{Pis'ma Zh.~Eksp.~Teor.~Fiz.~{\bf#1} #2 (#3) [JETP Lett.
{\bf#1} #4 (#3)]}
\def \pl#1#2#3{Phys.~Lett.~{\bf#1}, #2 (#3)}
\def \plb#1#2#3{Phys.~Lett.~B{\bf#1}, #2 (#3)}
\def \pra#1#2#3{Phys.~Rev.~A {\bf#1} #2 (#3)}
\def \prd#1#2#3{Phys.~Rev.~D {\bf#1}, #2 (#3)}
\def \prl#1#2#3{Phys.~Rev.~Lett.~{\bf#1}, #2 (#3)}
\def \prp#1#2#3{Phys.~Rep.~{\bf#1} #2 (#3)}
\def \ptp#1#2#3{Prog.~Theor.~Phys.~{\bf#1}, #2 (#3)}
\def \rmp#1#2#3{Rev.~Mod.~Phys.~{\bf#1} #2 (#3)}
\def \rp#1{~~~~~\ldots\ldots{\rm rp~}{#1}~~~~~}
\def \si90{25th International Conference on High Energy Physics, Singapore,
Aug. 2-8, 1990}
\def \slc87{{\it Proceedings of the Salt Lake City Meeting} (Division of
Particles and Fields, American Physical Society, Salt Lake City, Utah, 1987),
ed.~by C. DeTar and J. S. Ball (World Scientific, Singapore, 1987)}
\def \slac89{{\it Proceedings of the XIVth International Symposium on
Lepton and Photon Interactions,} Stanford, California, 1989, edited by M.
Riordan (World Scientific, Singapore, 1990)}
\def \smass82{{\it Proceedings of the 1982 DPF Summer Study on Elementary
Particle Physics and Future Facilities}, Snowmass, Colorado, edited by R.
Donaldson, R. Gustafson, and F. Paige (World Scientific, Singapore, 1982)}
\def \smass90{{\it Research Directions for the Decade} (Proceedings of the
1990 Summer Study on High Energy Physics, June 25 -- July 13, Snowmass,
Colorado), edited by E. L. Berger (World Scientific, Singapore, 1992)}
\def \stone{{\it B Decays}, edited by S. Stone (World Scientific, Singapore,
1994)}
\def \tasi90{{\it Testing the Standard Model} (Proceedings of the 1990
Theoretical Advanced Study Institute in Elementary Particle Physics, Boulder,
Colorado, 3--27 June, 1990), edited by M. Cveti\v{c} and P. Langacker
(World Scientific, Singapore, 1991)}
\def \yaf#1#2#3#4{Yad.~Fiz.~{\bf#1} #2 (#3) [Sov.~J.~Nucl.~Phys.~{\bf #1} #4
(#3)]}
\def \zhetf#1#2#3#4#5#6{Zh.~Eksp.~Teor.~Fiz.~{\bf #1} #2 (#3) [Sov.~Phys. -
JETP {\bf #4} #5 (#6)]}
\def \zpc#1#2#3{Z.~Phys.~C {\bf#1}, #2  (#3)}

\newpage

\centerline{\bf FIGURE CAPTIONS}
\bigskip

\noindent
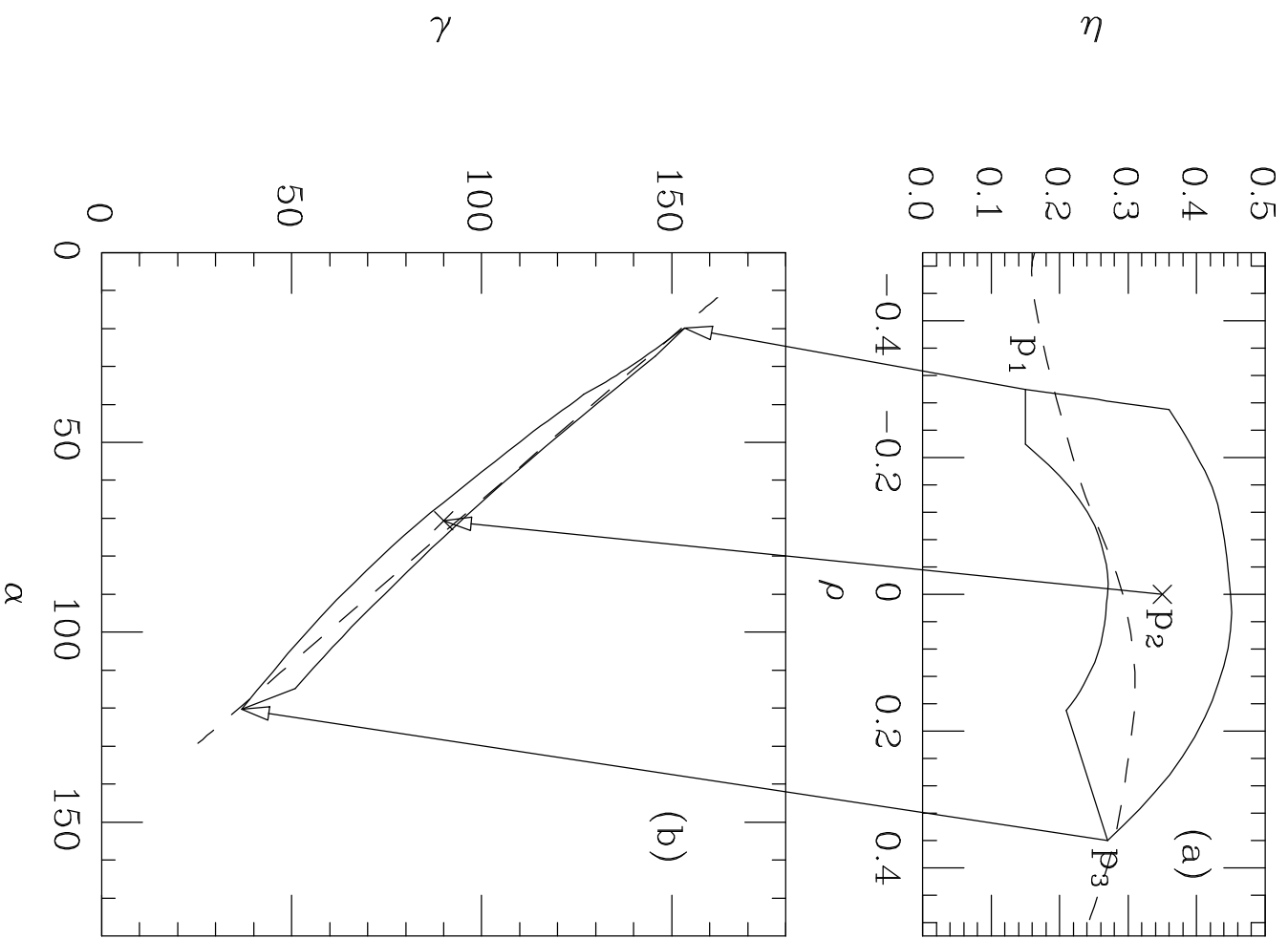
FIG.~1.
Allowed ranges of CKM parameters (bounded by solid lines). The points $p_1$,
$p_2$, and $p_3$ are described in Table I. (a) $(\rho,\eta)$ plane; (b) as
function of angles $\alpha$ and $\gamma$ of the unitarity triangle. A linear
least-squares fit to $p_1$, $p_2$, and $p_3$ yields $\gamma = 175^{\circ} -
1.16 \alpha$, as shown by the dashed straight line in (b); its map into the
$(\rho,\eta)$ plane is the dashed curve in (a). 
\bigskip

\noindent
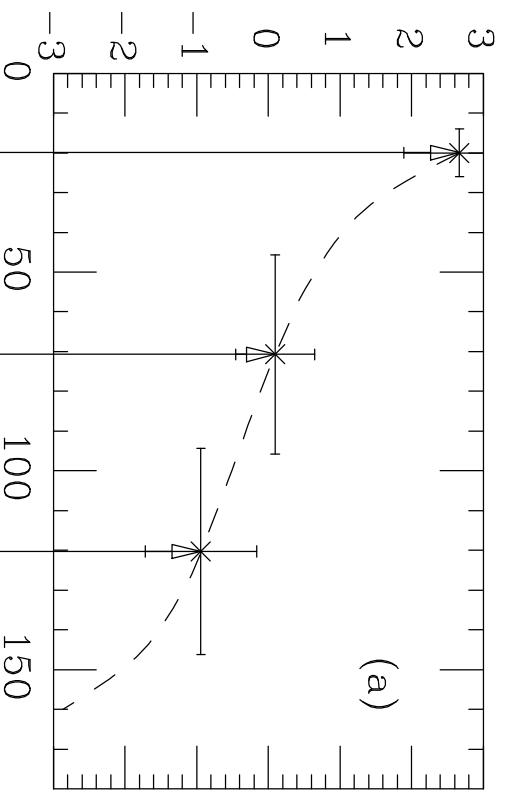
FIG.~2.
(a) The function $\cos(1.2 \alpha)/\sin \alpha$ as a function of $\alpha$,
and errors on $\alpha$ expected for a sample corresponding to 100 $B^0$
or $\overline{B}^0 \to \pi^+ \pi^-$ decays.  The three points, from left
to right, correspond to $p_1$, $p_2$, and $p_3$ of Table I, and $\delta = 0$
is assumed.  (b) Corresponding regions in $\alpha - \gamma$ plane.
\bigskip

\noindent
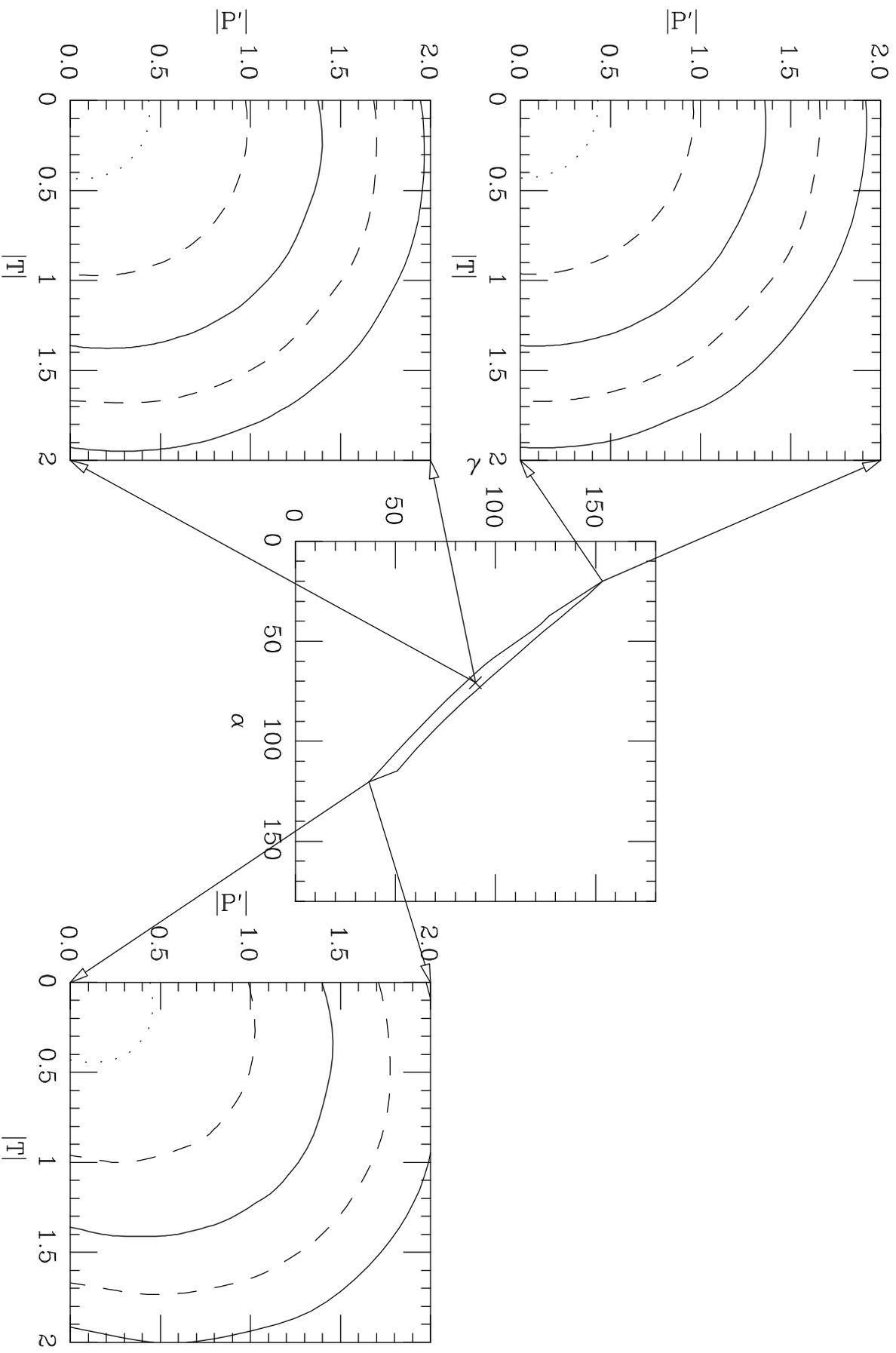
FIG.~3.
Contours in the $|T| - |P'|$ plane of fixed $S = A + D$ (sum of average $\pi^+
\pi^-$ and $\pi^\pm K^\mp$ branching ratio in units of $10^{-5}$), for
$\delta=0$.  Dotted curves:  $S = 0.2$.  Other curves, outward from origin: $S
= 1,~2,~3,~4$. 
\bigskip

\noindent
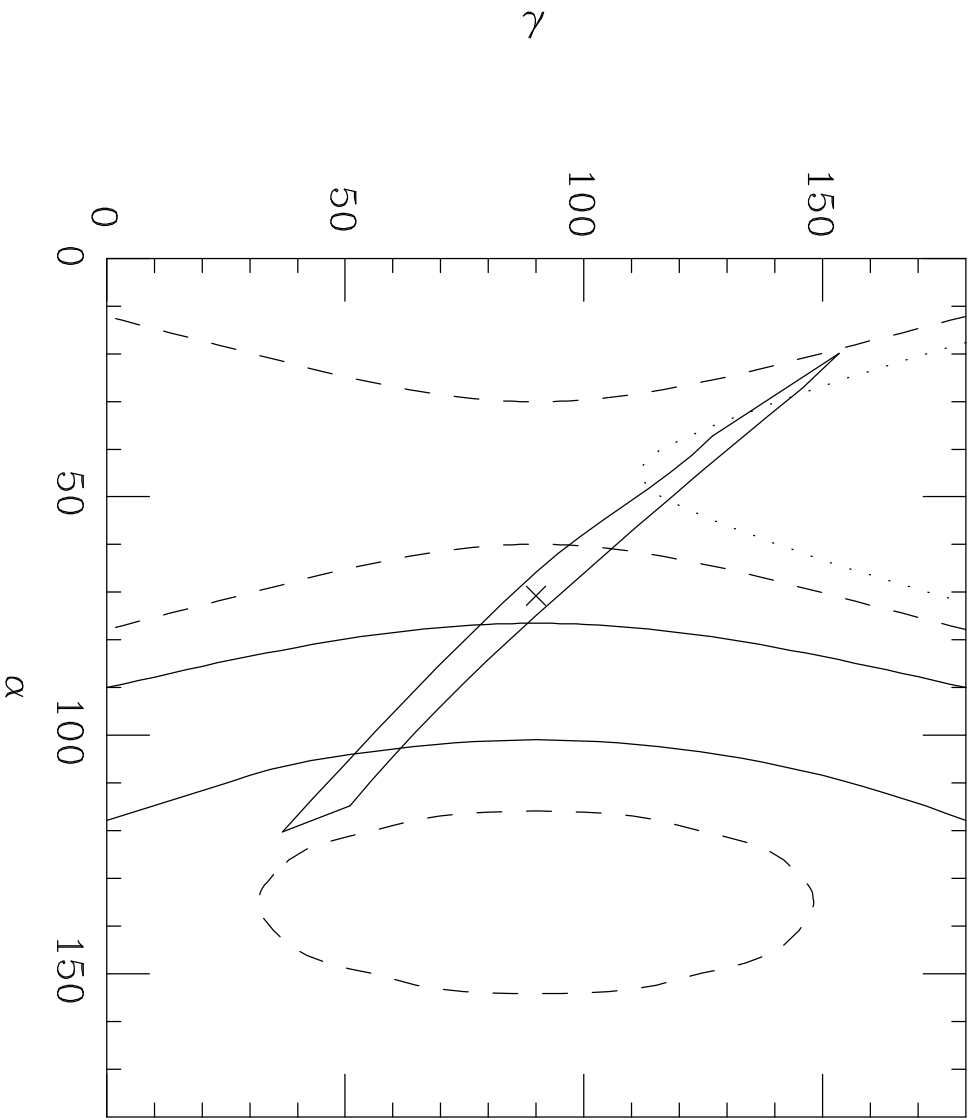
FIG.~4.
Contours of fixed $C$ for $|T| = |P'| = 1.1$ as functions of $\alpha$ and
$\gamma$, for $\delta=0$.  Only those branches intersecting the allowed region
are shown. Dotted curve: $C = -0.7$; dashed hyperbola: $C = -0.5$; solid
curves, from left to right: $C = 0,~1$; dashed ellipse: $C = 1.5$. 
\bigskip

\noindent
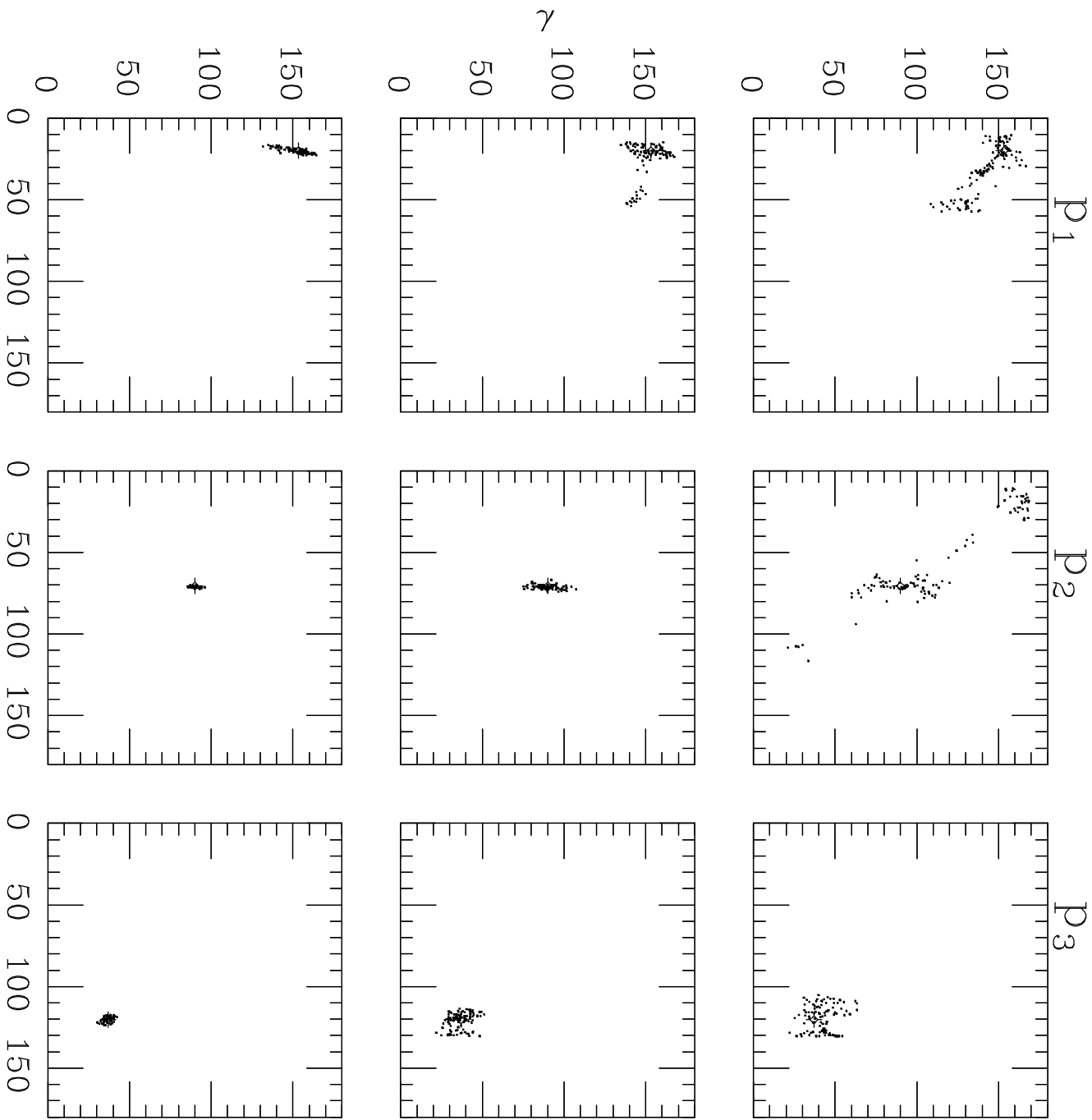
FIG.~5.
Scatter plots in the $\alpha - \gamma$ plane of 100 events generated according
to errors appropriate to samples of $M = 100,~1000,~10000$ events for the
points $p_1,~p_2,~p_3$ of Table I, for $\delta=0$.  Here the $\alpha$ and
$\gamma$ axes are plotted in degrees. 
\bigskip

\noindent
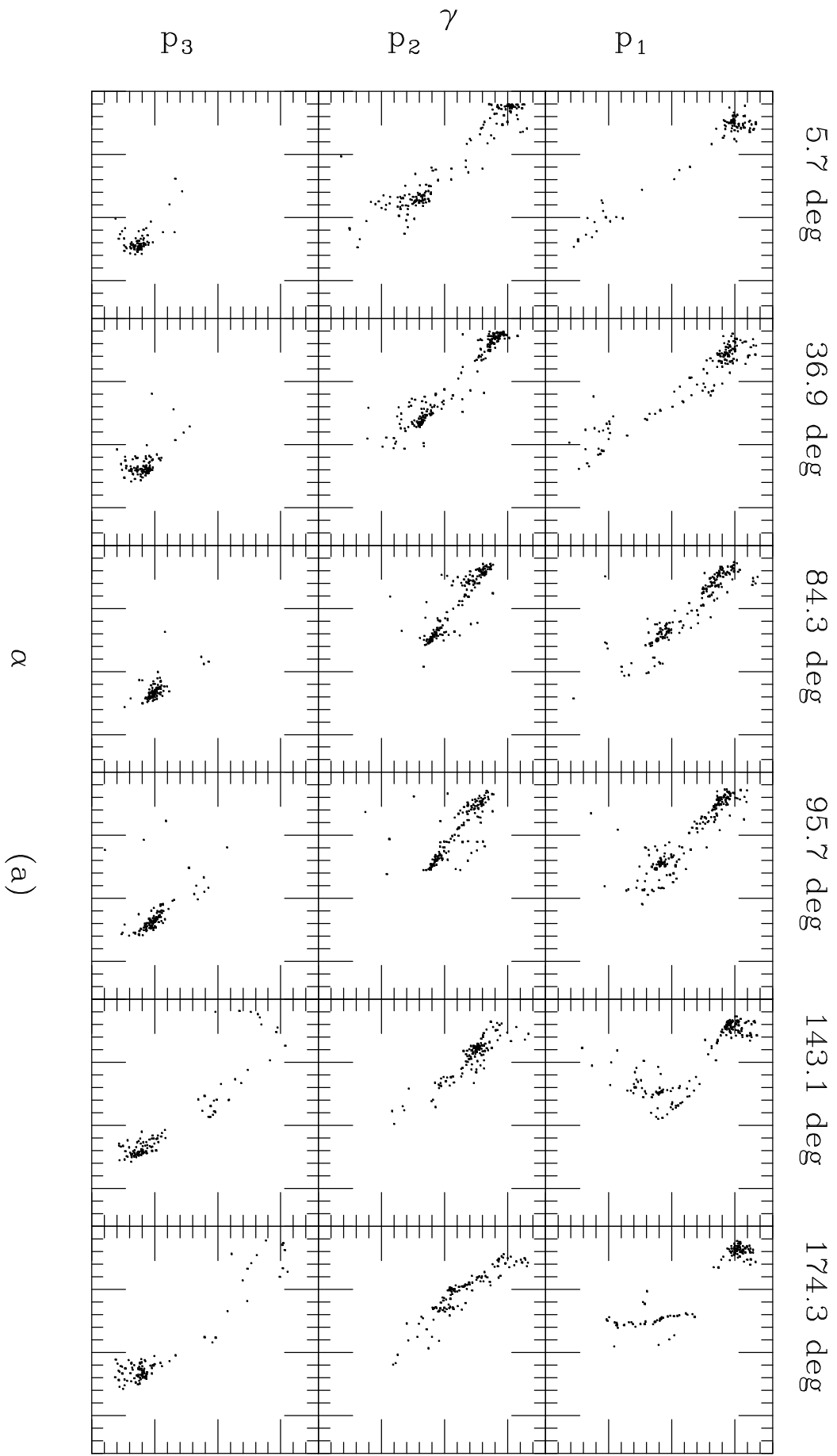
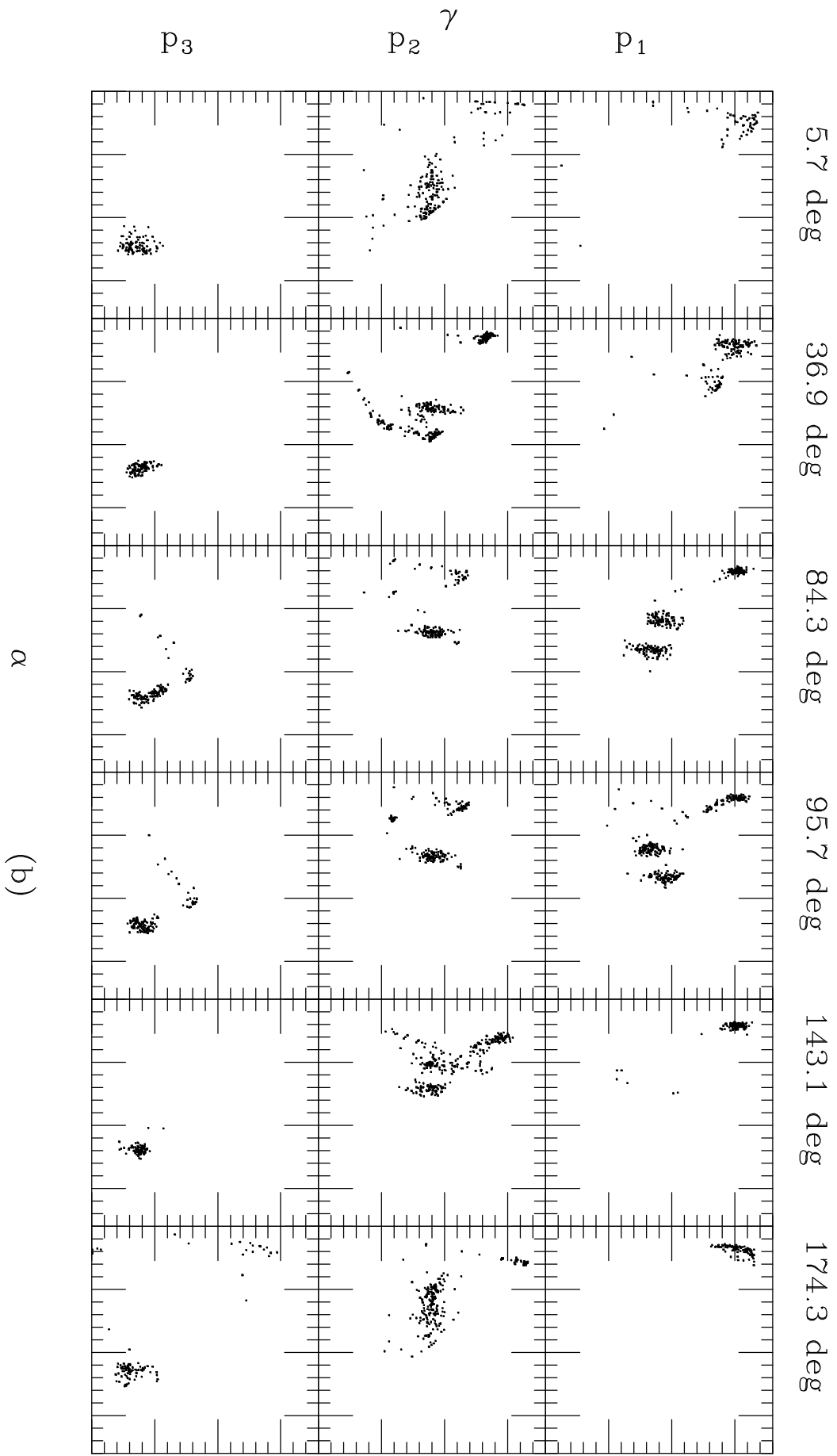
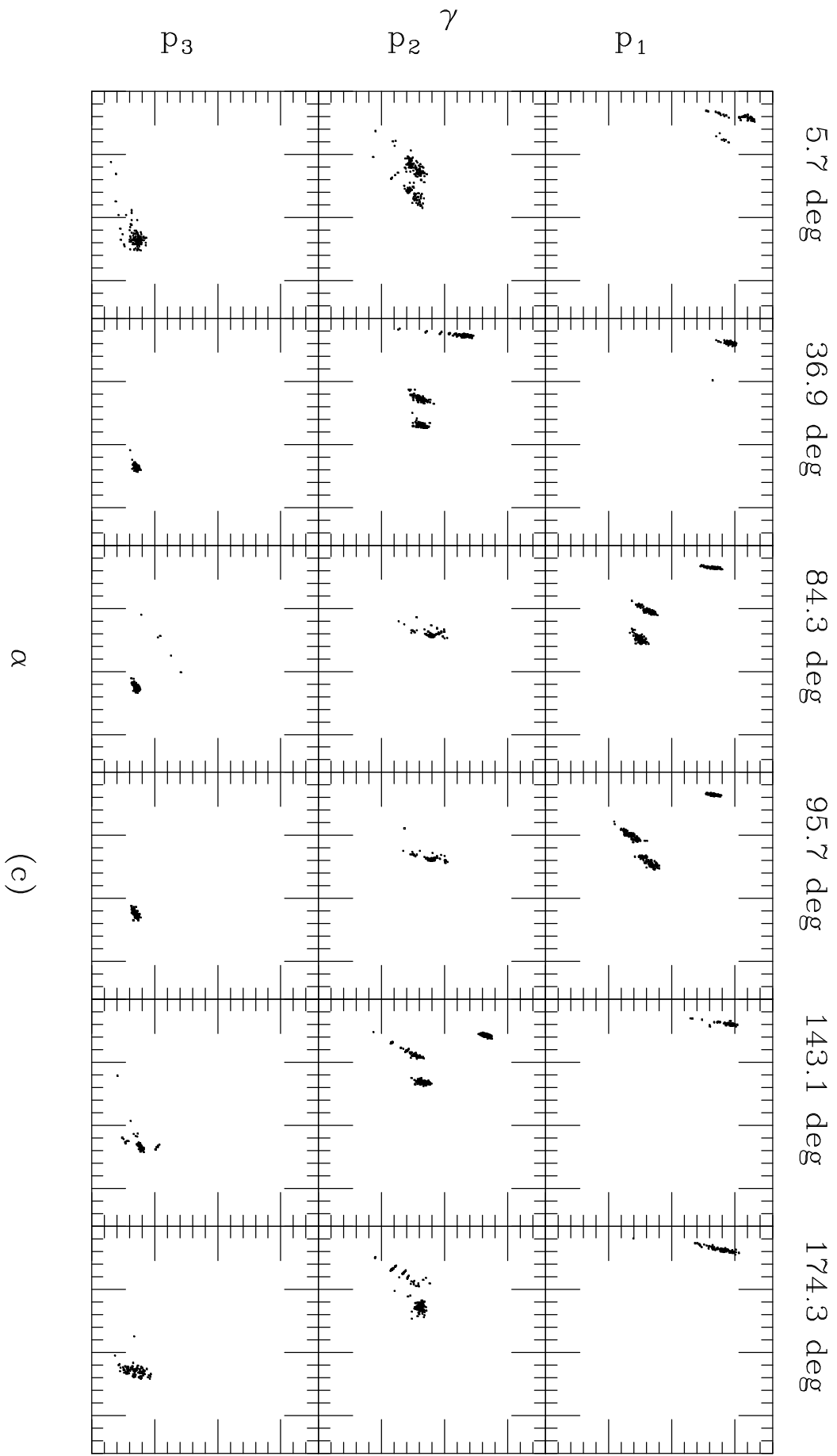
FIG.~6.
Scatter plots in the $\alpha - \gamma$ plane for nonzero input values of
$\delta$ (labels above columns) for the points $p_1,~p_2,~p_3$ of Table I
(labels to left of rows).  Here $0^\circ \le (\alpha,\gamma) \le 180^\circ$. 
(a) $M = 100$; (b) $M = 1000$; (c) $M = 10000$. 
\bigskip

\noindent
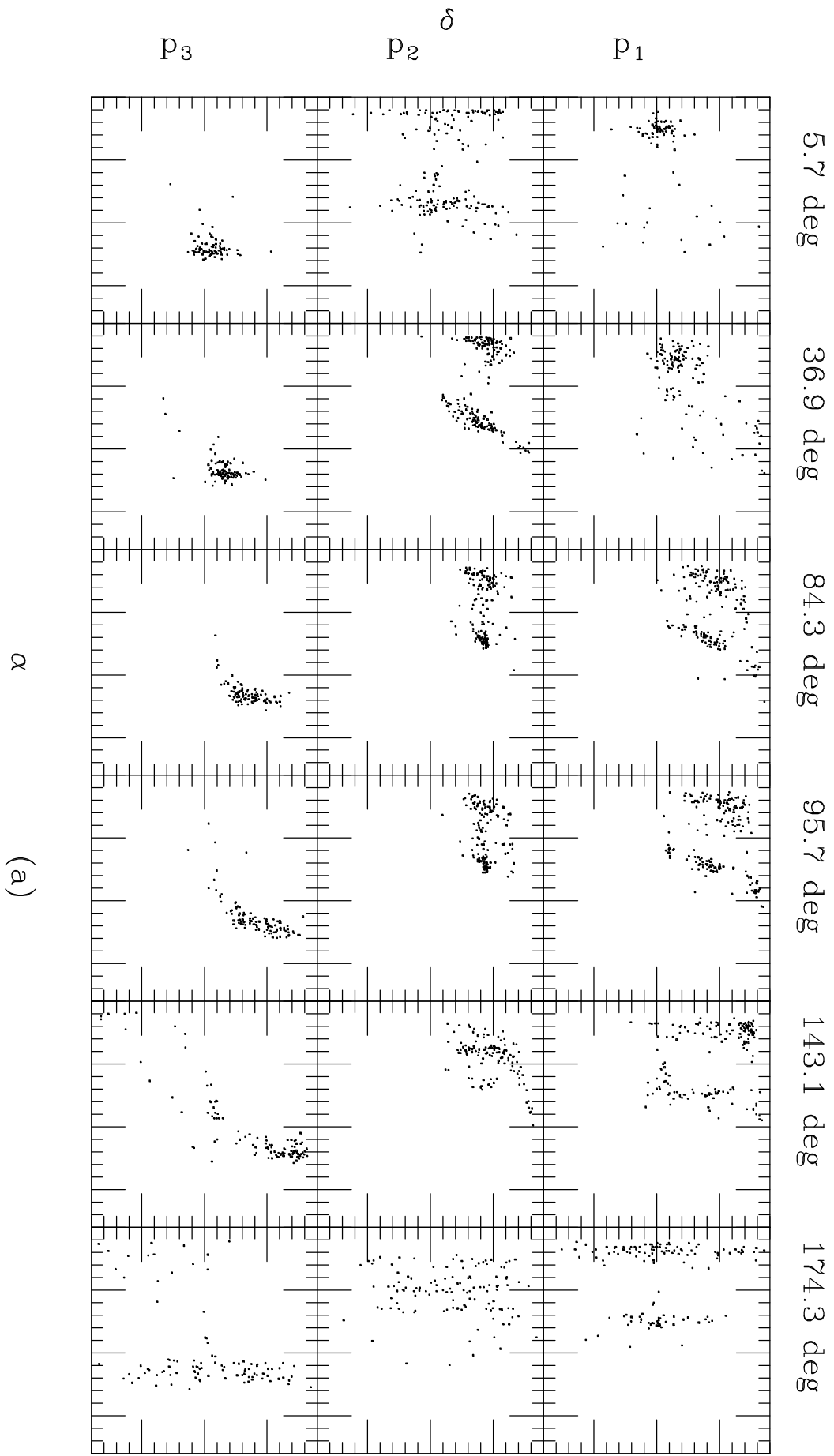
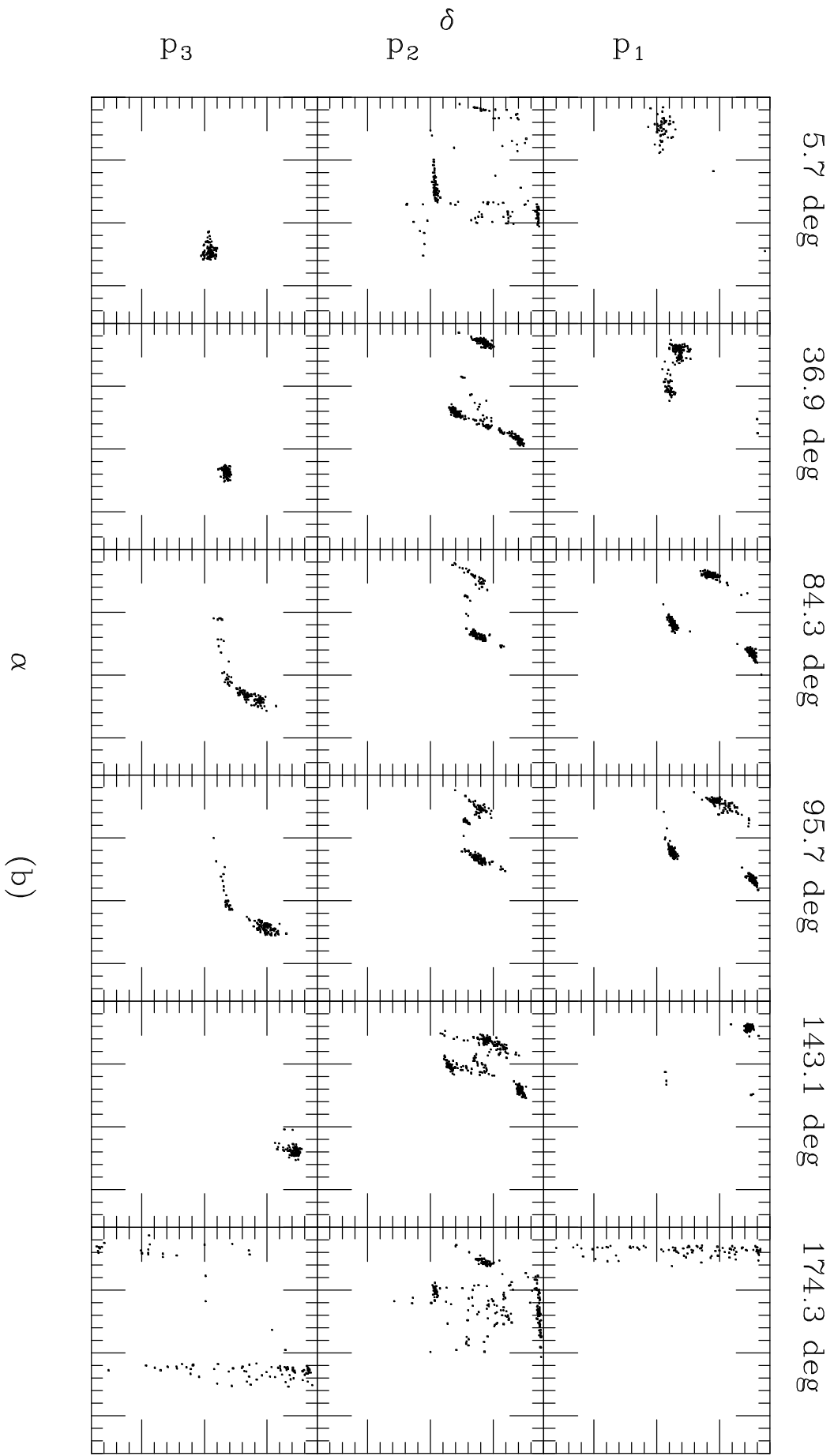
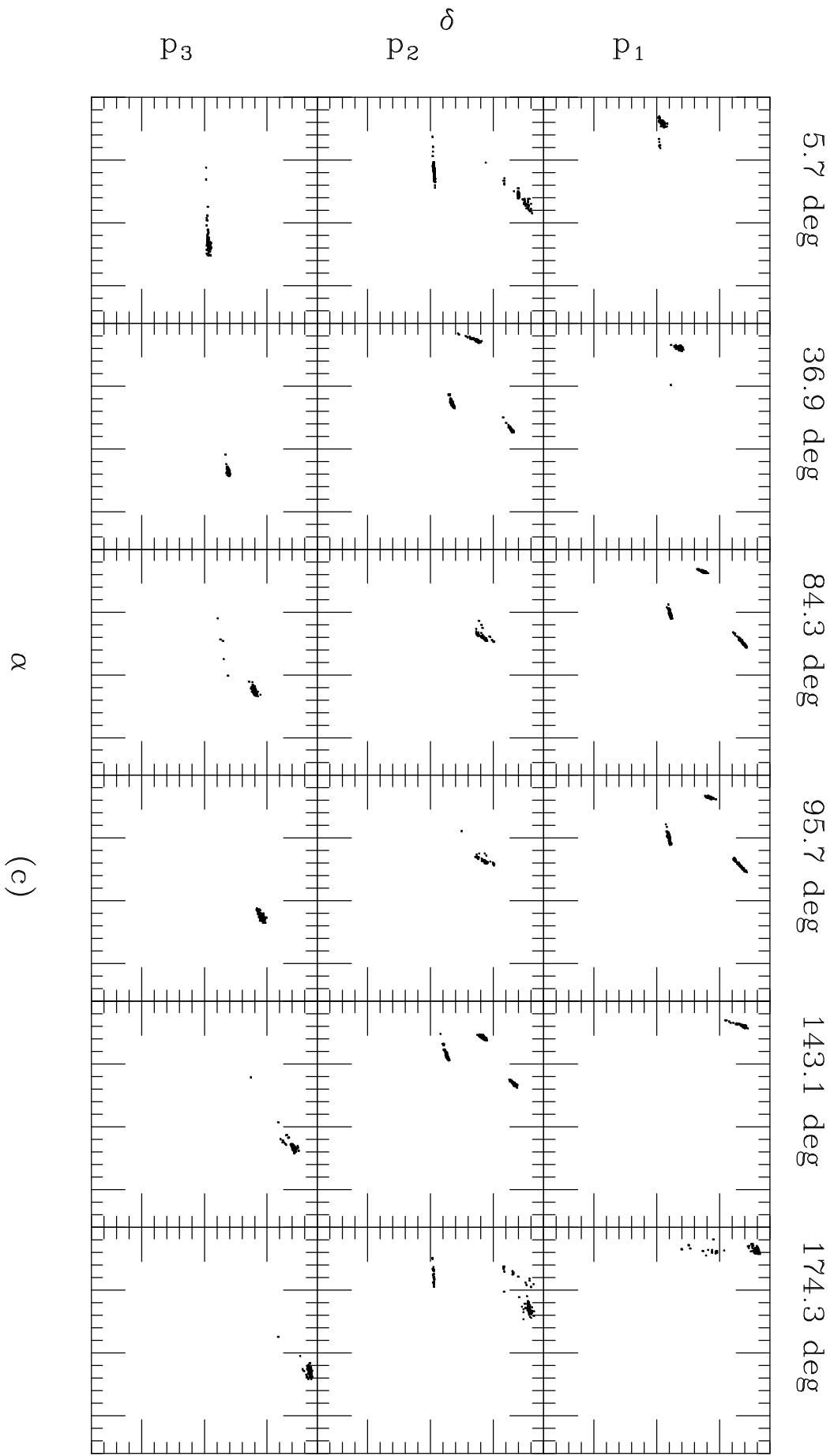
FIG.~7.
Scatter plots in the $\alpha - \delta$ plane for nonzero input values of
$\delta$ (labels above columns) for the points $p_1,~p_2,~p_3$ of Table I
(labels to left of rows).  Here $0^\circ \le \alpha \le 180^\circ$; $-180^\circ
\le \delta \le 180^\circ$.  (a) $M = 100$; (b) $M = 1000$; (c) $M = 10000$. 
\end{document}